\documentclass{wscpaperproc}
\usepackage{latexsym}
\usepackage{graphicx}
\usepackage{mathptmx}
\usepackage[T1]{fontenc}
\usepackage{mathtools}
\usepackage{listings}
\usepackage{braket}
\usepackage{verbatim}
\usepackage{wrapfig}
\usepackage{xcolor}
\usepackage{booktabs}

\usepackage{tikz}
\usetikzlibrary{arrows.meta, angles, quotes}

\usepackage{amsmath}
\usepackage{amsfonts}
\usepackage{amssymb}
\usepackage{amsbsy}
\usepackage{amsthm}
\usepackage{enumitem}

\usepackage{caption}

\usepackage[pdftex,colorlinks=true,urlcolor=blue,citecolor=black,anchorcolor=black,linkcolor=black]{hyperref}

\newtheoremstyle{wsc}
{3pt}
{3pt}
{}
{}
{\bf}
{}
{.5em}
{}

\theoremstyle{wsc}

\newcommand{\be}{\begin{equation}}
\newcommand{\ee}{\end{equation}}

\newcommand{\bE}{\mbox{\boldmath$E$}}

\def\eps{\varepsilon}

 \newcommand{\diff}{\mathrm{d}}

\newcommand{\Reals}{{\mathbb{R}}}

\newcommand{\Prob}{{\mathbb{P}}}
\newcommand{\Expect}{{\mathbb{E}}}

\def\bE{{\mathbb{E}}}

\begin{document}

%
%

\pagestyle{fancyplain}

\thispagestyle{plain}
\firstPageHead{}

\chead{\fancyplain{}{\itshape Blanchet, Squillante, Szegedy, and Wang}}

\rhead{}
\cfoot{}
\renewcommand{\headrulewidth}{0pt} 

\makeatletter
\let\@internalcite\cite
\def\cite{\def\@citeseppen{-1000}%
    \def\@cite##1##2{(##1\if@tempswa , ##2\fi)}%
    \def\citeauthoryear##1##2##3{##1 ##3}\@internalcite}
\def\citeNP{\def\@citeseppen{-1000}%
    \def\@cite##1##2{##1\if@tempswa , ##2\fi}%
    \def\citeauthoryear##1##2##3{##1 ##3}\@internalcite}
\def\citeN{\def\@citeseppen{-1000}%
    \def\@cite##1##2{##1\if@tempswa, ##2)\else{}\fi}%
    \def\citeauthoryear##1##2##3{##1 (##3)}\@citedata}
\def\citeA{\def\@citeseppen{-1000}%
    \def\@cite##1##2{(##1\if@tempswa , ##2\fi)}%
    \def\citeauthoryear##1##2##3{##1}\@internalcite}
\def\citeANP{\def\@citeseppen{-1000}%
    \def\@cite##1##2{##1\if@tempswa , ##2\fi}%
    \def\citeauthoryear##1##2##3{##1}\@internalcite}
\def\shortcite{\def\@citeseppen{-1000}%
    \def\@cite##1##2{(##1\if@tempswa , ##2\fi)}%
    \def\citeauthoryear##1##2##3{##2 ##3}\@internalcite}
\def\shortciteNP{\def\@citeseppen{-1000}%
    \def\@cite##1##2{##1\if@tempswa , ##2\fi}%
    \def\citeauthoryear##1##2##3{##2 ##3}\@internalcite}
\def\shortciteN{\def\@citeseppen{-1000}%
    \def\@cite##1##2{##1\if@tempswa, ##2\else{}\fi}%
    \def\citeauthoryear##1##2##3{##2 (##3)}\@citedata}
\def\shortciteA{\def\@citeseppen{-1000}%
    \def\@cite##1##2{(##1\if@tempswa , ##2\fi)}%
    \def\citeauthoryear##1##2##3{##2}\@internalcite}
\def\shortciteANP{\def\@citeseppen{-1000}%
    \def\@cite##1##2{##1\if@tempswa , ##2\fi}%
    \def\citeauthoryear##1##2##3{##2}\@internalcite}
\def\citeyear{\def\@citeseppen{-1000}%
    \def\@cite##1##2{(##1\if@tempswa , ##2\fi)}%
    \def\citeauthoryear##1##2##3{##3}\@citedata}
\def\citeyearNP{\def\@citeseppen{-1000}%
    \def\@cite##1##2{##1\if@tempswa , ##2\fi}%
    \def\citeauthoryear##1##2##3{##3}\@citedata}
%
%
%
\def\@citedata{%
    \@ifnextchar [{\@tempswatrue\@citedatax}%
                  {\@tempswafalse\@citedatax[]}%
}

\def\@citedatax[#1]#2{%
\if@filesw\immediate\write\@auxout{\string\citation{#2}}\fi%
  \def\@citea{}\@cite{\@for\@citeb:=#2\do%
    {\@citea\def\@citea{, }\@ifundefined
       {b@\@citeb}{{\bf ?}%
       \@warning{Citation `\@citeb' on page \thepage \space undefined}}%
{\csname b@\@citeb\endcsname}}}{#1}}%

%
\def\@citex[#1]#2{%
\if@filesw\immediate\write\@auxout{\string\citation{#2}}\fi%
  \def\@citea{}\@cite{\@for\@citeb:=#2\do%
    {\@citea\def\@citea{; }\@ifundefined
       {b@\@citeb}{{\bf ?}%
       \@warning{Citation `\@citeb' on page \thepage \space undefined}}%
{\csname b@\@citeb\endcsname}}}{#1}}%

%
\def\@biblabel#1{}
\makeatother



\newdimen\bibindent
\bibindent=0.0em
\def\thebibliography#1{\section*{\refname}\list
   {}{\settowidth\labelwidth{[#1]}
   \leftmargin\parindent
   \itemindent -\parindent
   \listparindent \itemindent
   \itemsep 0pt
   \parsep 0pt}
   \def\newblock{}
   \sloppy
   \sfcode`\.=1000\relax}


\setlength{\baselineskip}{12.7pt}

\title{Connecting Quantum Computing with Classical Stochastic Simulation}

\author{\begin{center}Jose Blanchet\textsuperscript{1}, Mark S.\ Squillante\textsuperscript{2}, Mario Szegedy\textsuperscript{3}, and Guanyang Wang\textsuperscript{4}\\
[11pt]
\textsuperscript{1}Dept.\ of Mgmt. Science and Eng., Stanford University, Stanford, CA, USA\\
\textsuperscript{2}Mathematics of Computation, IBM Research, Yorktown Heights, NY, USA\\
\textsuperscript{3}Dept.\ of Computer Science, Rutgers University, Piscataway, NJ, USA\\
\textsuperscript{4}Dept.\ of Statistics, Rutgers University, Piscataway, NJ, USA\end{center}
}

\maketitle

\vspace{-12pt}
\begin{abstract}
This tutorial paper introduces quantum approaches to Monte Carlo computation with
applications in computational finance. 
We outline the basics of quantum computing using Grover's algorithm for unstructured search to build intuition. We then move slowly to amplitude estimation problems and applications to counting and Monte Carlo integration, again using Grover-type iterations. A hands-on Python/Qiskit implementation illustrates these concepts applied to finance. The paper concludes with a discussion on current challenges in scaling quantum simulation techniques.
\end{abstract}

\section{Setting the Stage}
\label{sec:setting-stage}
\citeN{benioff1980} proposed a quantum version of a Turing machine and showed that such a device could be described using the equations of quantum mechanics. This was followed by the First Conference on the Physics of Computation, jointly organized by MIT and IBM in the summer of 1981, at which both Benioff and Richard Feynman gave keynote talks on quantum computing. Feynman’s keynote, in particular, addressed simulating physics with computers and introduced the concept of a ``quantum simulator''---a machine that, unlike a universal Turing machine, specializes in simulating quantum mechanical phenomena. This talk, subsequently published in~\cite{feynman1982}, is often credited with launching quantum computing as a distinct discipline. It ended with his now‑famous remark:

\smallskip 

\noindent{\em
``…nature isn’t classical, dammit, and if you want to make a simulation of nature, you’d better make it quantum mechanical, and by golly it’s a wonderful problem, because it doesn’t look so easy.'' -- Feynman}

\smallskip

Quantum computing has since evolved far beyond the simulation of quantum systems alone.

\medskip

\textbf{Why Quantum Computing?} Quantum computing has captured wide attention due to its potential to solve certain problems much faster than classical computers. For example, Shor's algorithm showed that a quantum computer could factor large numbers in polynomial time~\cite{Sho94,Sho97}, threatening classical RSA encryption which underpins many systems in the financial industry.
While practical applications are still emerging, the promise of exponential or significant speedups is a key motivator.

\textbf{Overview of Quantum Computing Architectures:} Depending on how we classify them, quantum computing hardware falls into two or three primary paradigms:
\begin{itemize}[noitemsep]
  \item \textbf{Traditional Gate-based quantum computers:} 
  Gate‑based quantum computers use logical quantum gates, quantum analogs of classical logic gates, to run algorithms. Platforms include superconducting quantum bits (e.g., IBM, Google, Rigetti), trapped‑ion systems (e.g., IonQ, Quantinuum), cold‑atom devices (e.g., QuEra), and photonic devices 
(e.g., Xanadu, PsiQuantum). These machines are universal, capable in principle of executing any algorithm in the class BQP, the complexity class for quantum polynomial‑time computation.
  \item \textbf{Quantum annealers (adiabatic quantum computers):} These are designed to solve optimization problems by exploiting quantum physics to find low-energy states. Notably, D-Wave's machines use quantum annealing to find the ground state of an Ising model or equivalently solve a Quadratic Unconstrained Binary Optimization (QUBO) problem~\cite{albash2018adiabatic}.
  However, the computational power of annealers remains an open research question.
\end{itemize}

While annealing machines work with a large number of qubits ($>10^3$) and potentially deliver at least quadratic speedups with additional significant constant-factor advantages, 
their classification as full quantum computers remains contested.
Meanwhile, any quadratic speedup would require a very careful control of error-correction resources because a large error-correction overhead can negate such performance gains.
In contrast, experiments with existing gate-based quantum computers showed that artificial problems can be crafted with input-output functionalities that are not simulable with realistic classical resources. In spite of this advance,
these systems continue to grapple with decoherence and gate errors—challenges that define the current Noisy Intermediate-Scale Quantum (NISQ) era, characterized by devices with hundreds of quantum bits and gate-error rates between 0.1-1\%. The field is now actively pursuing technologies that transcend these NISQ-era limitations to realize the full potential of quantum computation, while quantum-bit counts continue to grow exponentially.
For instance, IBM alone exhibited a factor of $10$ increase between $2016$ and $2019$, and nearly a factor of $5$ increase between $2020$ and $2022$.

\medskip

\textbf{Quantum Error Correction vs.\ Quantum Hardware Approaches to Reduce Errors:}  
Quantum states are fragile, and today's quantum bits are error-prone. Quantum error correction (QEC) encodes a single logical quantum bit into multiple physical quantum bits, detecting and correcting errors to store information reliably. QEC is theoretically necessary for scalable, fault-tolerant quantum 
computing~\cite{nielsen2010quantum}
unless one could build a device with zero error. In fact, 
at least a year
passed between Shor's algorithm for ideal, error-free quantum bits and the mathematical proof that quantum mechanics, in principle, permits its implementation on imperfect physical quantum bits.

The first step was made by Peter Shor himself, who constructed the first quantum error-correcting code (QECC), the 9-quantum-bit Shor code (independently,
a 7-quantum-bit code by Steane was discovered); refer to~\citeN{Shor1995,Steane1996}. Unlike classical codes, QECCs must correct both bit-flip and phase-flip errors and their combinations. However, a fully fault-tolerant architecture requires not only QECCs but also fault-tolerant gate constructions. To this end, researchers have focused on a subclass of QECCs, called CSS codes named after Calderbank, Shor, and Steane~\cite{Calderbank1996,Steane1996}, which allows separate correction of bit- and phase-flip errors and conveniently supports transversal Clifford gates. Virtually all practical QECCs in current use are CSS codes, or more generally stabilizer codes, an elegant mathematical formalism developed by \citeN{Gottesman1997}; see also \shortciteN{Calderbank1998}. The celebrated threshold theorem~\shortcite{AharonovBenOr1997,Kitaev1997,KnillLaflammeZurek1998} established that, if physical error rates lie below a constant threshold, arbitrarily long computations can be realized with only polylogarithmic overhead in quantum-bit count. At first, due to its complexity, it remained only a proof of concept.

Subsequent development included Kitaev's surface code~\cite{Kitaev2003} and its several variations
as practical QECCs. 
\citeN{BravyiKitaev2005}
introduced magic-state distillation,
greatly simplifying the 
result of Aharonov and Ben-Or. 
Yet
QEC
continues to raise deep questions both
in theory and practice, and it is a hot topic.

\smallskip

\textbf{QEC 
overhead.} Depending on quantum-bit quality, error correction can increase physical quantum-bit counts by factors of thousands and add significant time overhead. 
Estimates for tasks like breaking RSA with Shor’s algorithm often require millions of physical quantum bits. Even solving a chemistry problem, which could 
be the first killer application of quantum computing,
might theoretically need $100$ logical quantum bits, but if each logical quantum bit needs $\sim1,000$ physical quantum bits, that is $100,000$ physical quantum bits -- far beyond the few hundred available today.
Consequently, hardware researchers pursue 
three main strategies each of which have made recent advances:

\smallskip

1. \textbf{Improving coherence and control of quantum bits} across established platforms (superconducting circuits, trapped ions, neutral atoms, and so on).

\smallskip

2. \textbf{Clever ways of mitigating errors.} 
IBM has developed sophisticated error mitigation strategies \shortcite{zne_reference} including zero-noise extrapolation (ZNE) and readout error mitigation that reduce the impact of noise without full error correction. Additionally, emerging error correction techniques focus on detecting and correcting only the most frequent error types rather than achieving full fault tolerance, offering a practical intermediate approach for current hardware.

\smallskip

3. \textbf{Exploring novel architectures.} Topological quantum computing—pursued by Microsoft and several universities—aims to encode information in non‑Abelian anyons or other exotic states with intrinsic error protection. If realized, these quantum bits could exhibit dramatically lower error rates.
Nevertheless, formidable technical and materials challenges remain before topological quantum bits can become practical, if ever.

\smallskip

Incremental progress continues: the fidelity of quantum bits improve each year, and companies have roadmaps for devices with a thousand quantum bits, but significant scale-up is still required. Experts in the industry estimate a $5$-$20$ year timeframe for practically useful fault-tolerant quantum computers.

\textbf{Current Quantum-Solvable Problems (Focus on Quantum Versions of Monte Carlo):} What can quantum computers currently solve, or are expected to solve soon? We are currently in the NISQ
era, which means we must contend with the fact that quantum computing is inherently noisy—and in some approaches, even attempt to utilize this randomness constructively.

Algorithms like the Variational Quantum Eigensolver (VQE) due to \shortciteN{Peruzzo2014} and the Quantum Approximate Optimization Algorithm (QAOA) due to \shortciteN{Farhi2014} are heuristic hybrid approaches that combine quantum resources with classical feedback, aiming to achieve quantum advantage. However, despite these hybrid classical-quantum approaches, practical quantum advantage remains elusive. Nevertheless, theoretical results provide clear direction on several domains where quantum computing should excel once hardware capabilities mature sufficiently.

One such area is Monte Carlo simulation and integration, which is ubiquitous in science and finance. Quantum approaches to Monte Carlo methods aim to speed up the estimation of expected values, probabilities, and other statistical outputs. A prominent result by 
\citeN{montanaro2015quantum}
showed that quantum algorithms can quadratically speed up Monte Carlo methods under certain conditions. Specifically, if a classical Monte Carlo would take $N$ samples to achieve a certain error (uncertainty) $\epsilon$ (often $N\sim1/\epsilon^2$ for Monte Carlo), a quantum computer could in principle achieve a similar error with on the order of $\sqrt{N}$ samples (scaling as $1/\epsilon$) – an almost quadratic speedup. 

For example, in finance, Monte Carlo is used to price derivatives or estimate risk metrics. 
Quantum Amplitude Estimation (QAE) has been proposed to price financial derivatives with fewer samples than classical Monte Carlo due to \shortciteN{Rebentrost2018}. 
IBM researchers
\citeN{woerner2019}
demonstrated how QAE can compute risk measures like Value at Risk with far fewer simulations than a classical approach.
In general, 
further advances in quantum approaches to Monte Carlo methods should be welcome
because many industries (e.g., insurance, banking, engineering) have expensive Monte Carlo workloads.

It is important to note, though, that these advantages of quantum approaches to Monte Carlo assume we have quantum hardware capable of running the required circuits with low error. In practice, current hardware is not there yet -- for instance, QAE algorithms require either deep circuits (with Quantum Fourier Transforms) or multiple circuit repetitions that are challenging on NISQ devices. Researchers are exploring NISQ-friendly variants (like iterative amplitude estimation, which avoids some deep quantum operations) to test these ideas on current machines. So while theoretical speedups exist for Monte Carlo and other problems, demonstrating them experimentally will likely require further advances in hardware or error mitigation.


\section{Basic Elements of Quantum Computing}
\label{sec:basic-quantum}
Before diving into the quantum speedups of Monte Carlo, let us review the basic elements of quantum computing. We will then apply these notions to explain Grover’s search algorithm, which plays a key role in the quantum speedup of Monte Carlo methods. We start with the definition of two key concepts: qubits and superpositions. Our presentation here will be brief, but we refer the reader to~\shortciteN{nielsen2010quantum,de2019quantum,LipReg21} for additional details on quantum computing.

We shall use standard quantum notation throughout this paper, which includes representing quantum states (which, as we shall explain, are simply vectors) with the \emph{bra-ket} notation introduced by Paul Dirac. More specifically, a quantum state is represented as a column vector $\ket{\cdot}$ where different variables inside the \emph{ket} indicate different quantum states; for example, $\ket{s}$ and $\ket{s'}$ represent the (column) vector states $s$ and $s'$ of a quantum system. The corresponding \emph{bra}, represented as $\bra{\cdot}$, is the conjugate transpose of the ket (and therefore a row vector); for example, $\bra{s}$ and $\bra{s'}$ represent the conjugate transpose of the vector states $s$ and $s'$ of a quantum system. Finally, $\braket{s|s'}$ represents the inner (scalar) product formed between the two quantum states. 

\textbf{Qubits and Superposition.}
The basic unit of information in classical computing is the binary bit which takes on values of either $0$ or $1$.
In quantum computing, the basic unit of information is the quantum bit, or \emph{qubit}, which can exist in a \emph{superposition} of the basis states $\ket{0} = (1,0)^\top$ and $\ket{1} = (0,1)^\top$ (i.e., can exist in the multiple basis states simultaneously) with complex probability amplitudes. A general (pure) qubit state is written as
\begin{equation*}
\ket{s} = \alpha\ket{0} + \beta\ket{1}\,,
\end{equation*}
where \(\alpha,\beta\in\mathbb{C}\) satisfy \(\lvert\alpha\rvert^2 + \lvert\beta\rvert^2 = 1\). Here \(\ket{0}\) and \(\ket{1}\) correspond to the classical bits \(0\) and \(1\).

Probability amplitudes in quantum mechanics can interfere constructively (amplification) or destructively (cancellation), as they can be positive, negative, or complex numbers. This quantum interference is key to quantum computational speedups. When measuring the qubit $\ket{s}$, it collapses to state $\ket{0}$ with probability $|\alpha|^2$ or to state $\ket{1}$ with probability $|\beta|^2$, adhering to the normalization condition $|\alpha|^2+|\beta|^2=1$.

More generally, a quantum state is represented as a unit vector in the space of multiple qubits.
An $n$-qubit state is a unit vector in $(\mathbb{C}^2)^{\otimes n}$, where $\otimes$ denotes the tensor product. The standard basis, $B_n$, of this space has $2^n$ elements: $B_n = \{\ket{s}\mid \; s\in\{0,1\}^n\}$.

For example, the standard basis for a $2$-qubit state has $2^2 = 4$ elements:
~$\ket{00} = \ket{0} \otimes \ket{0} = (1, 0, 0, 0)^\top$,\\
~$\ket{01} = \ket{0} \otimes \ket{1} = (0, 1, 0, 0)^\top$,~
~$\ket{10} = \ket{1} \otimes \ket{0} = (0, 0, 1, 0)^\top$,~
~$\ket{11} = \ket{1} \otimes \ket{1} = (0, 0, 0, 1)^\top$.

\textbf{Operations and Computation.}
Quantum computing involves operations on the quantum state of \( n \) qubits.  
The state is a vector in a \( 2^n \)-dimensional complex Hilbert space (the complex analog of a Euclidean space equipped with an inner product).
Quantum operations correspond to multiplying this state vector by $2^n\times 2^n$ unitary matrices.
These operations are represented in quantum circuits as quantum gates acting on qubits, with the circuit complexity of a quantum algorithm defined by the number of gates and the circuit depth.
The exponentially large Hilbert space in which quantum computations are performed creates the potential for significant speedup over classical algorithms executed on digital computers.

A valid quantum operation applied to a qubit state is a unitary linear transformation (or matrix). Recall that a matrix $A$ is unitary if $AA^\dagger=I$, where $A^\dagger=\bar{A}^\top$ (i.e., the transpose of the complex conjugates of each entry). Equivalently, a linear operator is unitary if it preserves lengths and inner products, so unitary operators can be understood as `quantum rotations'. Of course, $AA^\dagger=I$ implies that any such $A$ is reversible. As three important examples, for single qubits, we have the so-called Pauli-X, Pauli-Y and Pauli-Z operators respectively represented by the following three matrices
$$ X =\displaystyle  \begin{bmatrix} 0 & 1 \\ 1 & 0 \end{bmatrix}, \qquad Y =\displaystyle  \begin{bmatrix} 0 & -i \\ i & 0 \end{bmatrix}, \qquad Z =\displaystyle  \begin{bmatrix} 1 & 0 \\ 0 & -1 \end{bmatrix},$$ where $i=\sqrt{-1}$. They happen to be not only unitary, but also Hermitian (i.e., $A=A^\dagger$).

The matrices $X$, $Y$ and $Z$, together with the identity matrix $I$, form a basis for all  $2\times2$ Hermitian matrices. Every unitary transformation $U$ on a single qubit can be decomposed into products of rotation matrices of the form $\exp(-i\theta Y/2)$ and $\exp(-i\phi Z/2)$ for $\theta \in[0,\pi]$ and $\phi\in[0,2\pi]$, multiplied by a possible global phase factor:
$U = e^{i\alpha} \exp(-i\beta Z/2) \exp(-i\gamma Y/2) \exp(-i\delta Z/2) .$
For example, the Hadamard operator or gate, represented by the matrix $H=\displaystyle \frac{1}{\sqrt{2}} \begin{bmatrix} 1 & 1 \\ 1 & -1 \end{bmatrix}$, can be expressed as $H=
e^{-3i\pi/2}\exp(-iY\pi/4) \exp(-iZ\pi/2)$ and it satisfies
\[
\ket{s} := H\ket{0}=\frac{\ket{0}+\ket{1}}{\sqrt{2}}.
\]
For a qubit initially at state $\ket{0}$, after applying the Hadamard gate, the qubit now has a $50\%$ chance of being either $0$ or $1$ when measured. The state is neither $0$ nor $1$ until the measurement, but in a coherent superposition of both $0$ and $1$. 

The $k$-qubit quantum gates (typically $k=1,2$) are unitary operators acting on the $k$-qubit Hilbert space with dimension $2^k$. 
The reader is referred to \shortciteN{QuantumGates1995} for more details on elementary quantum gates.

A sequence of quantum gates forms a quantum circuit, analogous to a classical logic circuit.
A major aspect of quantum algorithm design is determining how exponentially large unitary matrices
can be decomposed into a polynomial-length sequence of gate operations, when possible.
The typical quantum algorithm has the following structure:
~ $\textit{Input Preparation} \rightarrow \textit{Quantum Operations} \rightarrow \textit{Measurement} 
$ \\
Given that measurement in quantum computing is probabilistic, quantum algorithms often require multiple runs to obtain reliable statistics, and therefore the number of required runs plays an important role in the overall computational complexity of a given quantum algorithm.
In addition, the number of gates and the depth of the circuit define the circuit complexity of the quantum algorithm, which characterizes its implementation cost on quantum hardware.

\textbf{Entanglement.}
An important property in the setting of a quantum state with multiple qubits is \emph{entanglement}, which refers to a form of quantum correlations between different qubits such that the measurements of the state become correlated in ways that do not have a classical counterpart. 
To see this, one can reason as follows. Pure states, such as the pair ($\ket{0}, \ket{1}$), correspond to an arbitrary orthonormal basis. Each such base can be interpreted as a measuring setting. Physically, a basis can correspond to the orientation of a polarizer.

As noted above, an $n$-qubit quantum state can be expressed as the tensor product of $n$ single-qubit states.
On the other hand, it is possible for the $n$ qubits to be in a quantum state that cannot be expressed as the tensor product of $n$ single-qubit states.
These states, which cannot be expressed as a tensor product of its single-qubit states, are called entangled states.
One simple well-known example
is the two-qubit quantum state given by $(\ket{00} + \ket{11})/\sqrt{2}$, which is an instance of
the
quantum states considered in the classical paper
by \shortciteN{EPR1935}.
We observe from this simple example that, if measurement of the first qubit reveals $\ket{0}$, then the two-qubit quantum state collapses to $\ket{00}$, and similarly for a measurement of $\ket{1}$ collapsing to $\ket{11}$.
Hence, measurement of the first qubit immediately reveals the value of the second qubit which has not been measured.
This property of entanglement is one of the most important properties in creating the possibility of quantum algorithms that are more powerful than their classical counterparts.

To better understand the property of entanglement and how it arises, consider an $n_1$-qubit state $\ket{s_1}$ which can be expressed by $2^{n_1}$ complex numbers whose squared moduli sum to $1$ and an $n_2$-qubit state $\ket{s_2}$ which can be expressed by $2^{n_2}$ complex numbers whose squared moduli sum to $1$.
The tensor product $\ket{s_1s_2}=\ket{s_1} \otimes \ket{s_2}$ of these two quantum states, while it could take $2^{n_1+n_2}$ outcomes when measured, it depends on $2^{n_1} + 2^{n_2}$ complex numbers. However, an arbitrary quantum state in the joint space of the $n_1$-qubit state $\ket{s_1}$ and the $n_2$-qubit state $\ket{s_2}$ can be expressed by $2^{n_1+n_2}$ complex numbers whose squared moduli sum to $1$. The state $\ket{s_1s_2}$ lives in a $2^{n_1+n_2}$-dimensional Hilbert space, where those quantum states that cannot be expressed as a tensor product of the two quantum states $\ket{s_1s_2}$ are entangled states.
Entanglement therefore makes it possible to create a complete $2^n$-dimensional Hilbert space in which to perform our computations while using only $n$ qubits.
The reader may wonder if in the end it is possible to define a classical probability model on the $2^n$-dimensional that is consistent with quantum measurements. The answer is \textit{no}. This is precisely the implication of the thought experiment due to Clauser, Horne, Shimony, and Holt, leading to the so-called CHSH inequality which is part of a larger set of inequalities first derived by Bell and thus also known as the Bell inequality; see, e.g., 
\citeN{nielsen2010quantum}.

\paragraph{Grover's Algorithm: A Basic Example – Unstructured Search.} 
One of the most important quantum algorithms is Grover’s search algorithm which tackles the problem of finding a specific marked element $\omega$ from among a collection of $N$ elements labeled $1, \ldots, N$~\cite{Grover1996,Grover1997a}. Each element is associated with an element of the chosen basis. 
Grover’s algorithm finds the marked element $\omega$ with high probability using $O(\sqrt{N})$ well-chosen operations (to be discussed). This running time is known to be of optimal order~\shortcite{bennettbound1997}, with even the constant factor being optimal~\cite{Zalka1999}, in stark contrast to the best possible classical algorithm, which requires at least $\Omega(N)$ operations in the worst case to find $\omega$ with high probability. Hence, the computational complexity of Grover’s algorithm provides a quadratic speedup over that of the best classical algorithm for solving the general search problem.

The algorithm begins with the uniform superposition of all basis states on $N$ qubits. This is done by applying the so-called Hadamard gate to each qubit. Call this state $s$, with the element of interest $\ket{\omega}$. We can always write
\[
\ket{s} = \frac{1}{\sqrt{N}} \sum_{x=1}^{N} \ket{x} = \frac{1}{\sqrt{N}} \ket{\omega} + \frac{\sqrt{N-1}}{\sqrt{N}} \ket{s'},
\]
where $\ket{s'} = (N-1)^{-1/2} \sum_{x\neq\omega} \ket{x}$ is orthogonal to $\ket{\omega}$
and it represents the uniform superposition over all the unmarked elements. Note that if one were to measure the state $\ket{s}$ immediately after initialization, then the probability of observing the marked element $\omega$ is given by $\left(1/\sqrt{N}\right)^2 = 1/N$, which matches the probability in a classical search.  

Grover's algorithm dramatically increases this probability after a sequence of quantum operations. The algorithm can be understood by drawing a two-dimensional picture on the plane generated by the states $\ket{s}$ and $\ket{\omega}$. We place the horizontal axis on the line generated by $\ket{\omega}$ as shown in Figure~\ref{fig:grover_illustration} below. The Grover operator $G$ is the composition of two linear operators, $O$ and $D$, the first of which computes the reflection in the $\ket{s}$-$\ket{\omega}$ plane with respect to the $\ket{\omega}$-axis. More precisely, 
\[
O = 2 \ket{\omega}\bra{\omega}-I.
\]
Note that if $z=\alpha\ket{\omega}+\beta\ket{s'}$ then $Oz=\alpha\ket{\omega}-\beta\ket{s'}$, which verifies the reflection property with respect to the $\ket{w}$ axis. Note that this is just a sign flip in $s'$. 

The second operator computes a reflection with respect to the axis generated by $\ket{s}$ and then it flips the sign of the resulting vector, namely
\[
D = -(2 \ket{s}\bra{s} - I).
\]
Both of these operators are valid unitary operators (and thus valid quantum operations). Grover's operator therefore takes the form \(G = DO\). 

Let $\theta \in [0,\pi/2]$ be the angle between $\ket{s}$ and $\ket{w}$.  Note that we have $\bra{s}\ket{\omega}=\cos(\theta)=1/N^{1/2}$. 
Upon defining $\phi=\pi/2-\theta,$ we obtain $\sin(\phi)=1/N^{1/2}$, which implies both that $\phi\approx1/N^{1/2}$ and that
$$ \ket{s} = \sin(\phi)\ket{\omega}+\cos(\phi)\ket{s'}.$$
Each application \(r=G\ket{s}\) results in a state $r$ that is rotated by an angle \(\pi-2\theta
=\pi-2(\pi/2-\phi)=2\phi\). Then, $G^m$ yields a state that is rotated by an angle of $2m\phi$ units and therefore $$ G^m\ket{s} = \sin(2m\phi)\ket{\omega}+\cos(2m\phi)\ket{s'}.$$
Consequently, we want the resulting angle, namely $2m\phi$, to be as close as possible to $\pi/2$ so that $\sin(2\pi m)^2\approx1$, which leads to $m\approx\pi N^{1/2}/4$ for $N$ large. A measurement of the resulting state should then yield $\omega$ with high probability; see \shortciteN{BrassardHoyer1997,brassard2002quantum}.

\begin{center}
  \includegraphics[width=0.25\textwidth]{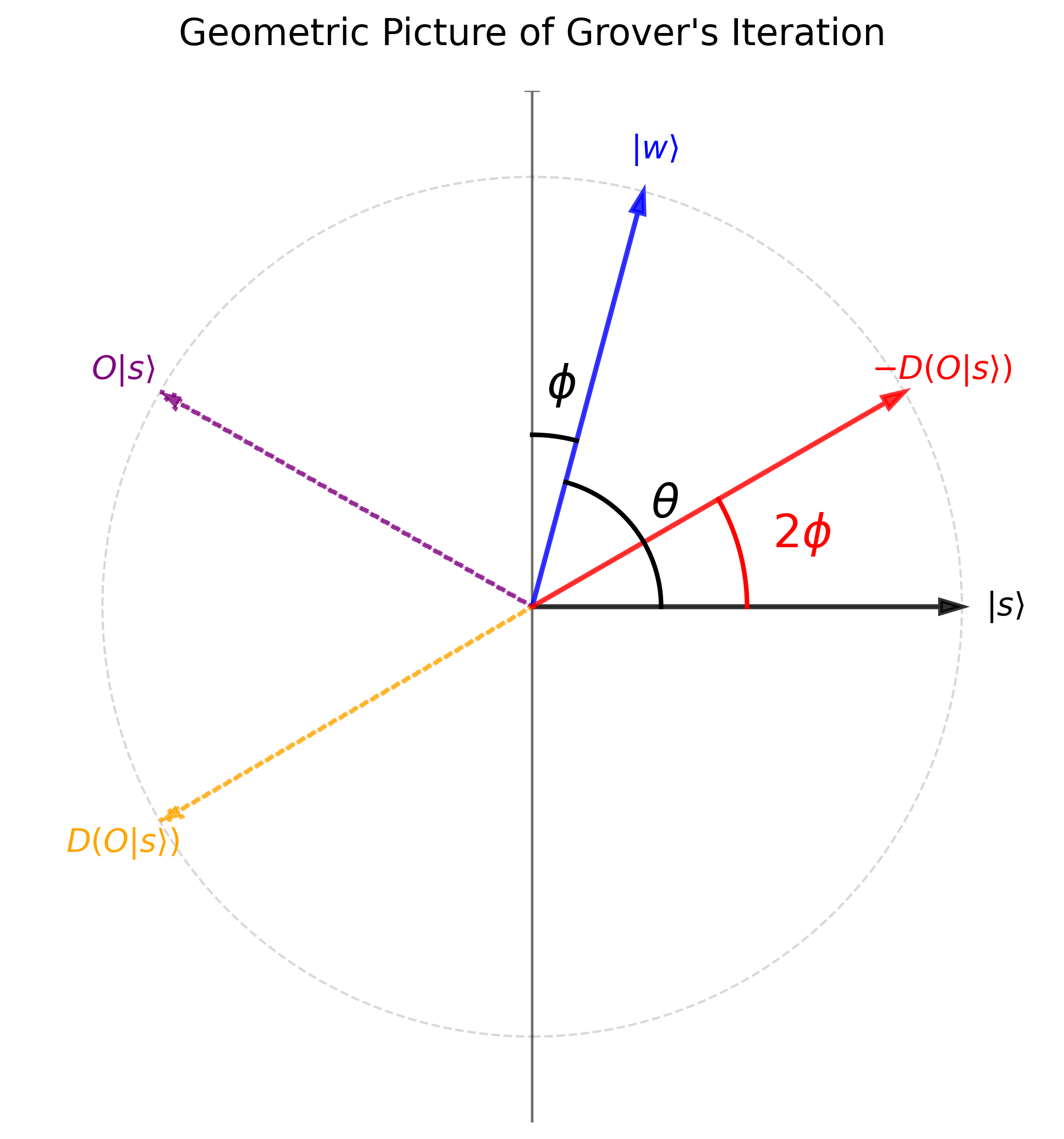}%
  \hspace{0.2\textwidth}%
  \includegraphics[width=0.25\textwidth]{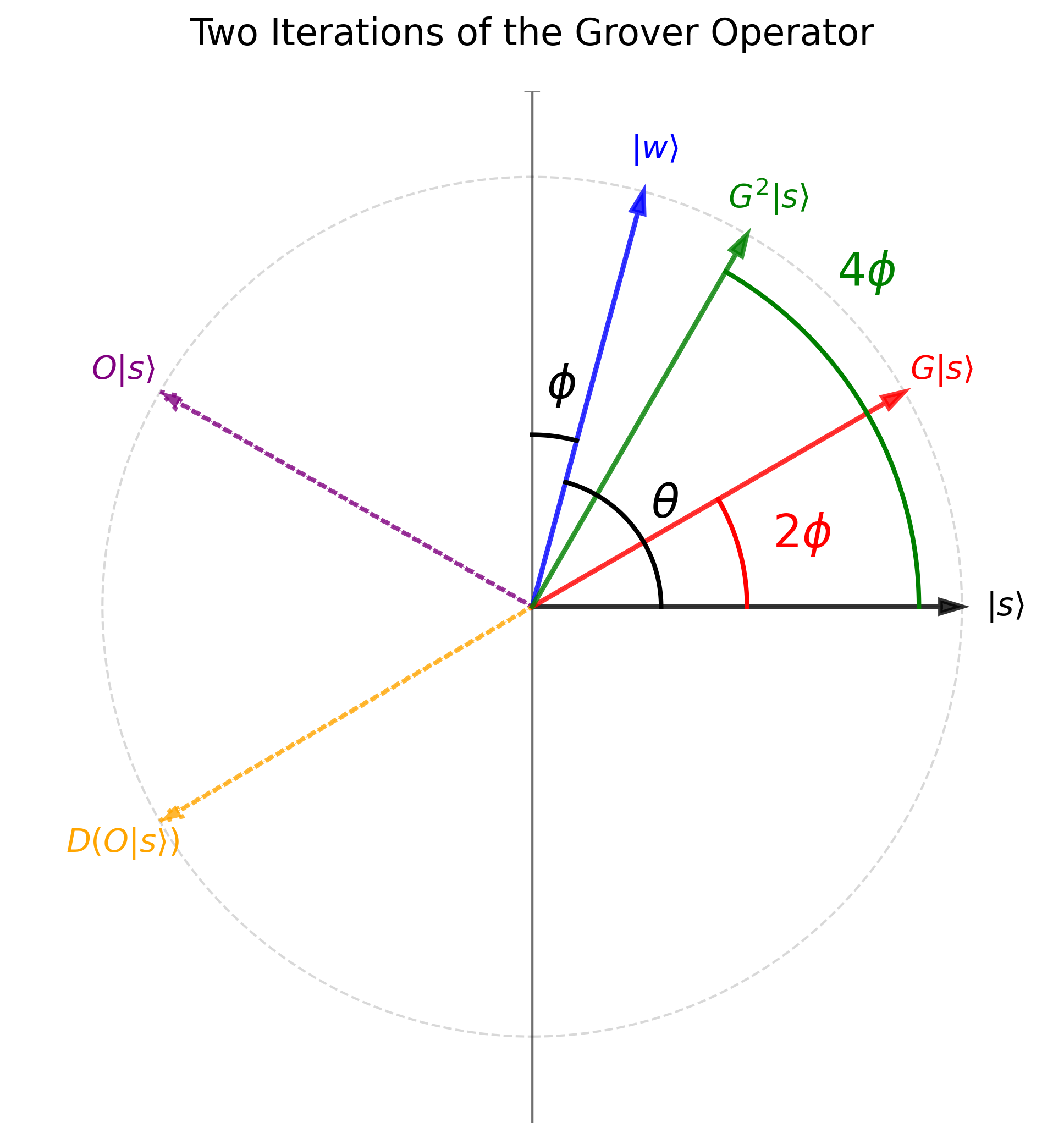}

  \vspace{0.5ex}
  \captionof{figure}{%
    Left: Geometric illustration of Grover’s iteration showing the relationship
    between the state vectors and angles $\theta$, $\phi$, and $2\phi$.
    Right: Geometric illustration of multiple iterations of Grover’s operator.
  }
  \label{fig:grover_illustration}
\end{center}


\section{From Grover
to Quantum {Speedup of} Monte Carlo {Methods}}
\label{sec:quantum-MC-Grover}

The central technical idea in Grover's algorithm is what is now known as \textit{amplitude amplification}, which uses quantum interference to boost the probability amplitude of a target state at a rate that exceeds what classical methods can achieve. Although originally introduced by Grover for unstructured search, this technique was later formalized and generalized by Brassard et al., making it applicable to a much wider class of computational problems. Following the path of \shortciteN{brassard2002quantum} and \citeN{montanaro2015quantum}, we will take a brief journey from Grover's search algorithm to quantum approaches to Monte Carlo methods.
Figure~\ref{fig2} (Left) 
shows a schematic diagram of this progression. The figure traces a progression of quantum algorithms. The horizontal axis shows how broad the underlying problem class is: a point farther to the right means you are answering a richer statistical query, moving from a binary search (Grover’s algorithm) to counting marked items, and finally to estimating an expectation as in Monte Carlo. The vertical axis captures the algorithmic sophistication required to obtain the quadratic speedup that these methods share: Grover’s search is the simplest routine, quantum counting builds on it, and the ``quantum speedup of Monte Carlo'' is the most general and technically involved. A summary table of the main algorithms is provided in Table \ref{tab:algorithms}.

\begin{center}
  \includegraphics[width=0.25\textwidth]{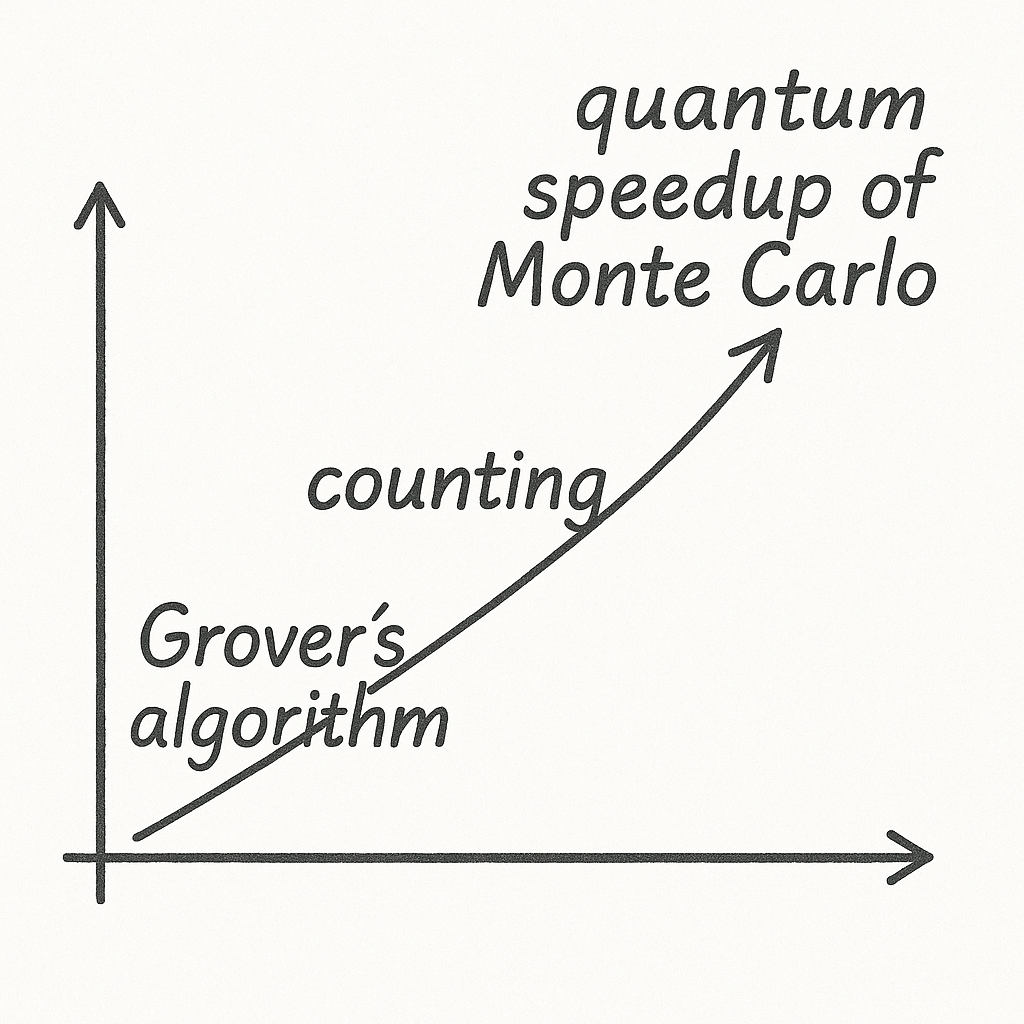}%
  \hspace{0.15\textwidth}%
  \includegraphics[width=0.4\textwidth]{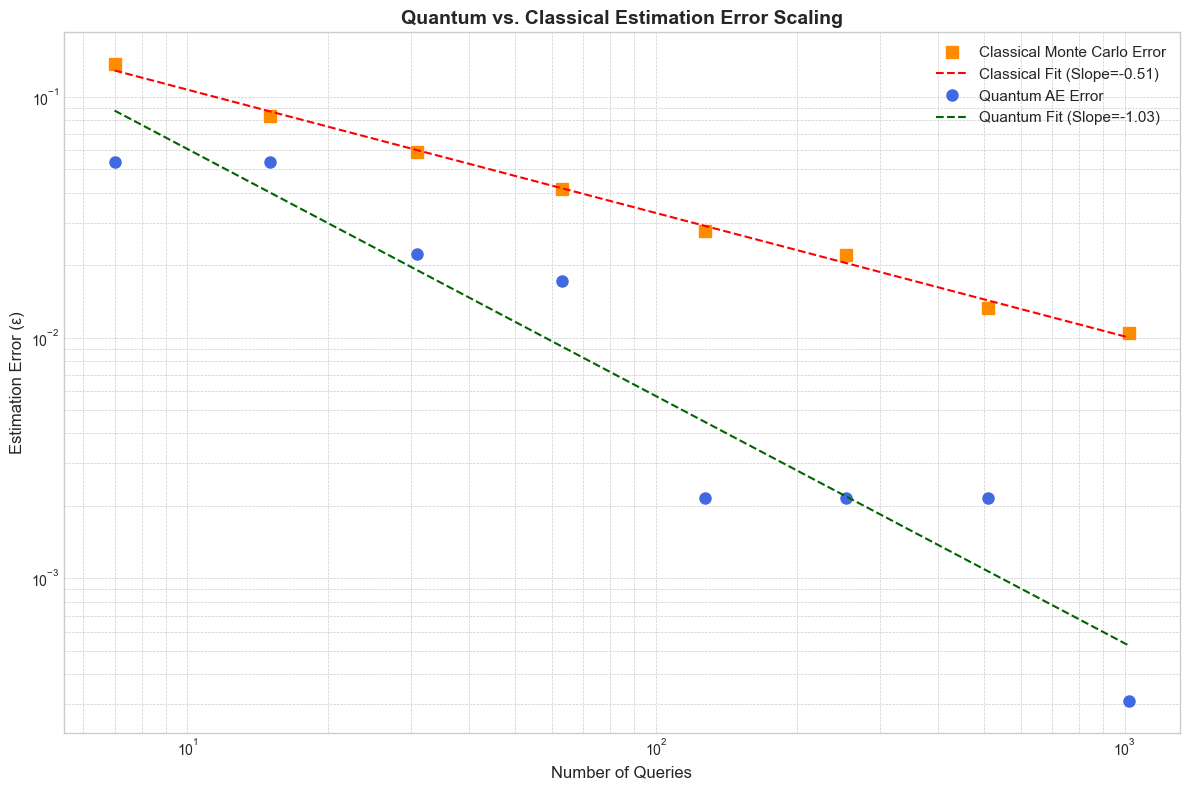}

  \vspace{0.5ex}
  \captionof{figure}{%
    Left: From Grover's algorithm to 
    quantum speedup of Monte
Carlo methods.
    Right: Illustration of quadratic speedup.
  }
  \label{fig2}
\end{center}

\begin{table}[h!]
\centering
\caption{Main algorithms and their classical analogs.}
\label{tab:algorithms}
\scalebox{0.8}{
\begin{tabular}{l l p{6.5cm}}
\toprule
\textbf{Quantum Algorithm} & \textbf{Classical Analog} & \textbf{Description of Uses} \\
\midrule
Grover's Algorithm & Unstructured Search & Finds an item in an unsorted collection. \\
\addlinespace
Quantum Amplitude Estimation (QAE) & Bernoulli Parameter Estimation & Estimates a probability  using Grover-style rotations to amplify the ``success'' state. \\
\addlinespace
Quantum-Accelerated Monte Carlo & General Monte Carlo Methods & Extends  QAE to estimate the expected value of a general function, $\mathbb{E}[f(X)]$. \\
\bottomrule
\end{tabular}}
\end{table}

\subsection{From Grover to Counting: A First Approach to Amplitude Estimation}

Consider the problem of estimating the number of elements $k$ that satisfy some property. This problem is equivalent to computing $
  p = \frac{k}{N},$
which we will do entirely by Grover rotations and simple measurements.  Recall from Section~\ref{sec:basic-quantum} that we set
$  \sin\varphi = \sqrt{\tfrac{k}{N}} = \sqrt p$  and  $
  \cos\varphi = \sqrt{1-p}.$

There are two approaches to this problem. One approach is to estimate the phase $\varphi$ and then estimate $p$, which is the approach taken by \shortciteN{brassard2002quantum}. The method we pursue here is based on the approach of \citeN{aaronson2020quantum}. Once again, we assume that the uniform superposition has been applied. Then, we assume that $k$ distinguished elements are marked in the basis; as before, they altogether are represented by $\omega$, which leads to
\[
  \ket{s}
  = \frac{1}{\sqrt{N}}\sum_{x=1}^N \ket{x}
  = \sin\varphi\ket{\omega} + \cos\varphi\ket{s'}.
\]
Since each Grover iterate \(G=DO\) rotates this $2$-dimensional plane by \(2\varphi\), we obtain just as in the Grover picture that $G^m\,\ket{s}
  = \sin\!\bigl((2m+1)\varphi\bigr)\,\ket{\omega}
  + \cos\!\bigl((2m+1)\varphi\bigr)\,\ket{s'}.$ Measuring $G^m\ket{s}$ in the basis $\ket{x}$, $x=1,\ldots,N$, thus yields the marked item \(\omega\) with probability
\[
  p_m \;=\;\Prob(\text{“marked”})
      \;=\;\sin^2\!\bigl((2m+1)\varphi\bigr).
\]
We can repeat this prepare-Grover-iterate-measure loop \(R\) times (obtaining $R$ i.i.d.\ replications) and therefore compute an empirical estimate
\(\hat p_m(R)= p_m  + \eta_{m,R}\), where $\eta_{m,R}$ represents the error captured by the Central Limit Theorem and thus $\eta_{m,R}=O_p(1/R^{1/2})$. We then invert the rotation classically, rendering
\[
  \widehat\varphi
  \;=\;\frac{1}{2m+1}\,\arcsin\!\bigl(\sqrt{p_m  + \eta_{m,R}}\bigr)
  \qquad \mbox{ and } \qquad
  \widehat p = \sin^2\!\widehat\varphi.
\]
We now can perform an expansion using elementary calculus to verify that, if $m=O(1/\epsilon)$ and $R$ grows slowly (say $\log(1/\epsilon)$), we can conclude that \(\lvert\widehat p - p\rvert\le\varepsilon\).  Hence, we recover the same quadratic speedup observed earlier directly as a consequence of Grover's algorithm and classical statistical measurements.

To sketch some of the details, dropping the subindex in $\eta_{m,R}$, let $\theta=(2m+1)\varphi$ where 
in contrast with Grover
we want to choose $\theta$ close to $\pi/4$. 
We aim to make \textit{small changes} in $p$ show up as \textit{big, easily measured} changes. Thus, we want to maximize $\frac{d}{d\theta}\sin^2\theta = \sin(2\theta)$, which leads to the target $\theta \approx \pi/4$. Then, 
$\arcsin\!\bigl(\sqrt{p_m + \eta}\bigr)
\approx \theta + \eta + O(\eta^2)$ and $\hat \varphi  = \varphi + O(\eta/m)$. From this we can propagate the error into $\hat p$ to conclude $|p - \hat p| = O(\eta /m)$, thus deducing that $m = O(1/\eps)$ and $R$ essentially constant does the trick.

Figure~\ref{fig2} (Right)
illustrates the quadratic speedup achieved by quantum amplitude estimation. We estimate the success parameter $p$ of a Bernoulli random variable using both quantum amplitude estimation (blue) and classical Monte Carlo (orange) via  Qiskit. We additionally fitted a regression line of $\log(\text{error})$ versus $\log(\text{cost})$ for both methods. The fitted curve clearly indicates that amplitude estimation achieves a $1/\epsilon$ scaling, offering a quadratic speedup compared to the $1/\epsilon^2$ scaling of classical Monte Carlo.

\subsection{Amplitude Estimation for a General State‐Preparation Operator}

Suppose we have a unitary \(A\) that prepares the state
$\ket{\psi} \;=\; A\ket{0}
  \;=\;\sqrt{p}\,\ket{\omega} \;+\;\sqrt{1-p}\,\ket{\psi'},$
  $p\in[0,1],$
where \(\ket{\omega}\) is the “good” state and \(\ket{\psi'}\) spans the orthogonal complement so that
\(\{\ket{\omega},\ket{\psi'}\}\) is an orthonormal basis of the two‐dimensional subspace of interest.
As before, we set $\sin\varphi = \sqrt{p}$
  and
  $\cos\varphi = \sqrt{1-p}.$
Define the two reflections as
\[
  O \;:=\; I \;-\; 2\,\ket{\omega}\!\bra{\omega}
  \qquad\qquad\qquad\qquad\quad \text{(oracle on }\ket{\omega}\text{),}
\]
\[
  D_\psi \;:=\; 2\,\ket{\psi}\!\bra{\psi} \;-\; I
  \;=\;
  A\,(2\,\ket{0}\!\bra{0} - I)\,A^\dagger
  \qquad\text{(diffusion about }\ket{\psi}\text{).}
\]
Their product
$G = D_\psi\,O$
rotates the \(\{\ket{\omega},\ket{\psi'}\}\) plane by \(2\varphi\).  In particular,
\[
  G^m\,\ket{\psi}
  = \sin\!\bigl((2m+1)\varphi\bigr)\,\ket{\omega}
  \;+\;\cos\!\bigl((2m+1)\varphi\bigr)\,\ket{\psi'}.
\]
A measurement in the \(\{\ket{\omega},\ket{\psi'}\}\) basis then yields a “good” state with probability $p_m =\sin^2\!\bigl((2m+1)\varphi\bigr).$
By repeating the prepare-rotate-measure sequence \(R\) times, one obtains an empirical
\(\hat p_m\approx p_m\).  Inverting classically then renders
\[
  \widehat\varphi
  = \frac{1}{2m+1}\,\arcsin\!\bigl(\sqrt{\hat p_m}\bigr)
  \qquad \mbox{ and } \qquad 
  \widehat p = \sin^2\!\widehat\varphi.
\]
Choosing \(m=O(1/\varepsilon)\) (and constant \(R\)) guarantees
\(\lvert\widehat p - p\rvert\le\varepsilon\) using \(O(m)=O(1/\varepsilon)\)
applications of \(G\), thus generalizing the quadratic‐speedup amplitude estimation
to any state‐preparation unitary \(A\).

\subsection{Quantum Monte Carlo via Grover-Style Amplitude Estimation}

Now we are ready to estimate
$
  \mathbb{E}[\,f(X)\,] \;=\; \sum_{x\in\{0,1\}^{Nd}} q(x)\,f(x)
  \;=\; p,
$
where \(f(x)\in[0,1]\). 
This can be seen as an approximation to the expectation of a Lipschitz $f$ function in $[0,1]^d$ assuming that $f(x) \in [0,1]$. 
We will apply a very similar procedure as above. 

\paragraph{1. State preparation operator \(\mathcal A\).}  
Define \(\mathcal A\) by
\[
  \mathcal A\ket{0^{Nd+1}}
  \;=\;
  \sum_{x}
    \sqrt{q(x)}\;\ket{x}_{Nd}
    \;\Bigl(\sqrt{f(x)}\,\ket{1}
           +\sqrt{1-f(x)}\,\ket{0}\Bigr)
  =: \ket{\psi},
\]
so that the final ancilla qubit has “\(\ket1\)” with overall probability
\(\;p = \Expect[f(X)]\).
Here, the notation $0^{Nd+1}$ indicates the dimension of the vector $0$ and the notation $\ket{x}_{Nd}$ indicates a vector in the $Nd$ quantum register.
Note that, if $q(x)$ is uniform, we can use the Hadamard gate as in previous examples. Moreover, if $f(x)$ can be evaluated easily on a classical computer, then the unitary $\mathcal A$ can be easily implemented as well.
A specific example is given in Section \ref{Subsec:Amplitud_in_Action}. 

\paragraph{2. Grover-geometry.}  
Set
\[
  \sin\varphi = \sqrt{p},
  \qquad
  \cos\varphi = \sqrt{1-p},
  \qquad
  \ket{\psi'}
  = \frac{1}{\sqrt{1-p}}\sum_{x}\sqrt{q(x)(1-f(x)}\,\ket{x}\ket0.
\]
Then, in the two-dimensional subspace spanned by
\(\ket{\omega}:=\sum_{ x}\sqrt{q( x)\,f( x)}\,\ket{x}\ket1\)
and \(\ket{\psi'}\), we have
\(\ket{\psi}=\sin\varphi\,\ket{\omega}+\cos\varphi\,\ket{\psi'}\).

\paragraph{3. Reflections and rotation.}  
Define the “good” oracle as $O := I - 2\,\ket{\omega}\!\bra{\omega},$
and the diffusion about \(\ket\psi\) as
\[
  D_\psi := 2\,\ket{\psi}\!\bra{\psi} - I
         = A\,(2\,\ket0\!\bra0 - I)\,A^\dagger.
\]
Then their product $G = D_\psi\,O$
rotates that 2-dimensional plane by \(2\varphi\):
\[
  G^m\ket{\psi}
  = \sin\!\bigl((2m+1)\varphi\bigr)\,\ket{\omega}
  + \cos\!\bigl((2m+1)\varphi\bigr)\,\ket{\psi'}.
\]

\paragraph{4. Re-prepare \& measure.}  
For each run, we have
\[
  \ket0\xrightarrow{\mathcal A}\ket\psi
  \xrightarrow{G^m}G^m\ket\psi
  \xrightarrow{\text{measure ancilla}}
  \;\;\text{“good” with probability }p_m
  = \sin^2\!\bigl((2m+1)\varphi\bigr).
\]
Repeating this \(R=O(1)\) times yields \(\hat p_m\approx p_m\), from which we 
obtain $\widehat\varphi
  = \frac{1}{2m+1}\,\arcsin\!\sqrt{\hat p_m}$
  and
  $\widehat p = \sin^2\!\widehat\varphi.$
Choosing \(m=O(1/\varepsilon)\) renders \(\lvert\widehat p-p\rvert\le\varepsilon\)
with only \(O(1/\varepsilon)\) total calls to \(G\), i.e., a quadratic speedup.

\citeN{montanaro2015quantum} extends QAE to the general Monte Carlo problem. Assuming $f$ has finite second moment (with a known upper bound) under $p$, Montanaro's algorithm proceeds by applying the idea to dyadic intervals. Precisely, it suffices to assume that $f\ge 0$ (you can write $f=f^+-f^-$). Then, simply estimate using the above algorithm  $\Expect [f(X)/2^{l+1}I(f(X) \in [2^l,2^{l+1}])]$ for $l = O(\log(1/\eps))$. This strategy achieves $\tilde O(1/\epsilon)$ cost to estimate $\Expect[f(X)]$ with an error up to $\epsilon$.
Here, using standard notation, $\tilde{O}(\cdot)$ is the same as $O(\cdot)$ but simply ignoring logarithmic factors.

\subsection{Further Developments}
There have been many further developments since  \citeN{montanaro2015quantum}, some of which we mention here.

\textbf{Algorithm improvements:} Montanaro's algorithm has two limitations. First, it requires a predetermined upper bound on the variance. Second, the error bound includes additional polylogarithmic factors in the parameter $\epsilon$. Both limitations are resolved in the recent work of \citeN{hamoudi2021quantum} and \citeN{kothari2023mean}. In particular, Kothari and O'Donnell's algorithm is state-of-the-art. Consider a random variable \(\mathbf{y}\) with a finite mean \(\mu\) and finite variance \(\sigma^2\). Suppose a user wants to estimate \(\mu\) and has access to the code that generates samples of \(\mathbf{y}\). Kothari and O'Donnell's algorithm requires \(O(n)\) executions of this code and produces an estimator \(\hat{\mu}\) that, with high probability, satisfies
$\lvert \hat{\mu} - \Expect[\mathbf{y}] \rvert \leq \sigma/n.$

The mean estimation problem  has been generalized in other directions. \shortciteN{cornelissen2022near} consider the multivariate mean estimation problem and present near-optimal quantum algorithms that explicitly depend on the accuracy level \(\epsilon\), the dimensionality \(d\), and the covariance matrix \(\Sigma\). When the random variable has infinite variance but a finite \((1+\delta)\)-th moment, \shortciteN{blanchet2024quadratic} propose a quantum algorithm similar to \citeN{montanaro2015quantum} that also achieves near-quadratic speedup.

\textbf{Practical algorithms on near-term devices:}
Nearly all quantum approaches to Monte Carlo methods rely on 
QAE.
The original QAE method employs a quantum phase estimation subroutine due to \shortciteN{brassard2002quantum} that achieves optimal scaling in theory. However, it requires many controlled amplification operations followed by a quantum Fourier transform, resulting in a circuit that is both deep and resource-intensive. Consequently,  this approach is often impractical for near-term quantum devices. 

There have been several new algorithms that aim to perform QAE without phase estimation. 
\shortciteN{suzuki2020amplitude}
employ a maximum likelihood estimator (MLE) to estimate the mean from the combined measurement data. Similarly, \shortciteN{grinko2021iterative} introduce an iterative algorithm where MLE is applied at each iteration to update the estimate. Both algorithms  seem to achieve better empirical performance than the original QAE. Meanwhile, \citeN{aaronson2020quantum} adopt a sequence of hypothesis tests, based on amplitude amplification, to progressively narrow down the possible range of the true amplitude.


\section{Financial Example Using Quantum Approach to Monte Carlo}
\label{sec:financial-example}

Quantum versions of Monte Carlo offer a promising approach for financial applications, such as pricing basket options whose payoffs depend on high-dimensional correlated assets.  Classical Monte Carlo can suffer from slow convergence in such settings. The quadratic speedup of quantum approaches to Monte Carlo makes it possible to achieve comparable accuracy with far fewer samples.

Many problems in finance involve computing a quantity of the form $\Expect_{Q}[U(X)]$, where $U$ represents a utility or payoff function. The random element $X$ has distribution $Q$, and $X$ may be a random variable or a random vector, or a continuous-time stochastic process (i.e., a path). In this section, we illustrate the concepts through a series of examples related to option pricing, with increasing levels of complexity. In the subsequent section, we present a case study on applying quantum algorithms to Credit Risk Analysis using Qiskit. For more extensive discussions, we refer the reader to survey papers such as those by \shortciteN{orus2019quantum} and \shortciteN{herman2023quantum}.

\paragraph{Analytically solvable:} Consider a simple European call option, where the underlying asset price \( S_t \) at time \( t \) evolves as \( S_t = S_0 \exp(\sigma W_t + (r - \sigma^2/2)t) \), with \( W_t \) denoting a Brownian motion and \( \sigma, r \) being fixed constants. The payoff function is defined by \( U(x) := \max\{0, x - K\} \) for some fixed strike price \( K \), and the option price is defined as \( \mathbb{E}[U(S_T)] \). In this setting, the Black–Scholes formula provides a closed-form expression for the option price.
\paragraph{`Vanilla' Monte Carlo:} When the payoff function is more complex, an analytic solution is often unavailable. For instance, the European power option has a payoff of the form  $U(x) = \max\{0, x^p - K\}.$ For general values of $p$, this expectation is typically estimated using standard Monte Carlo methods. In such cases, quantum approaches to Monte Carlo~\cite{montanaro2015quantum} can provide a quadratic speedup over classical Monte Carlo approaches.
    Similarly, quantum approaches to Monte Carlo can be directly used to speed up the pricing of basket options.
\paragraph{Multilevel Monte Carlo:} Sometimes the underlying process  also has to be approximated. For example, the Local Volatility model assumes the underlying price satisfies: $
        \diff S_t = \mu S_t \diff t + \sigma(S_t, t) S_t \diff W_t.$    
        In practice, the underlying process must be approximated using a numerical discretization method, such as the Euler–Maruyama scheme. In this setting, the estimation accuracy is influenced by both the discretization error (bias) and the number of Monte Carlo samples (variance). Classically, the multilevel Monte Carlo method~\cite{giles2015multilevel,rhee2015unbiased} is widely used to optimally balance bias and variance, allowing one to reach the desired level of precision with minimal computational cost.
    Similarly, quantum-accelerated multilevel Monte Carlo methods~\shortcite{an2021quantum} have been proposed to reduce the required number of samples. The degree of speedup ranges from quadratic to sub-quadratic, depending on the structure of the problem.
\paragraph{Monte Carlo inside Monte Carlo (Nested Monte Carlo):} 
Many financial derivatives, such as American options, permit early exercise. As a result, their valuation depends not only on the current underlying price but also on the expected future behavior. Formally, this leads to the problem of estimating a nested expectation: $
\mathbb{E}\left[g(X,\mathbb{E}[\phi(X,Y) \mid X])\right],$
where $X$ may represent the price tomorrow, and $Y$ represents the price in a week. The function $g$ captures the dependence on both the current situation and the anticipated future. Besides option pricing, this quantity arises in many other decision-making scenarios. Concrete examples include  Bayesian experimental design~\shortcite{goda2022unbiased}, and conditional stochastic optimization~\shortcite{hu2024contextual}. A variant of Multilevel Monte Carlo~\cite{blanchet2015unbiased,giles2019decision} achieves the optimal classical complexity of $O(1/\epsilon^2)$ under mild assumptions. In contrast, the quantum-inside-quantum algorithm introduced by \shortciteN{blanchet2025non} achieves the quantum-optimal complexity of $\tilde{O}(1/\epsilon)$.

\section{Case study: Credit Risk Analysis via Qiskit}
Now we illustrate how to exploit quantum approaches to Monte Carlo to analyze credit risk using the \textsc{Qiskit} package. The problem considered here is still simpler than real-world problems. Nevertheless, we will provide a representative examination of how these circuits are constructed and implemented, and how amplitude estimation is performed using current software. We follow the setting of \shortciteN{egger2020credit} and the online tutorial
\cite{QiskitEcosystem},
while also providing additional explanations.

\subsection{Problem Setup}
We start by considering the \textit{Gaussian conditional independence (GCI) model} of \citeN{rutkowski2015regulatory} which defines a joint distribution over $(Z, X_1, \ldots, X_K) \in \Reals \times \{0,1\}^K$ as follows.
Let $Z$ be a standard Gaussian random variable, and let $X_1, X_2, \ldots, X_K$ be binary-valued random variables that are conditionally independent given $Z$. The conditional distribution is then defined by $
 \Prob(X_k = 1 \mid Z = z) := F\left(\frac{F^{-1}(p_k^0) - \sqrt{\rho_k} z}{\sqrt{1 - \rho_k}}\right),$
where \( p_k^0 \in [0,1] \) and \( \rho_k \in (0,1) \) are given parameters, and \( F \) denotes the cumulative distribution function of the standard normal distribution. We denote $\Prob(X_k = 1 \mid Z = z)$ by $p_k(z)$, $k=1,\ldots,K$.

In financial applications, the random variable $X_k$ represents the event that obligor $k$ defaults within the risk measurement horizon. See also \citeN{rutkowski2015regulatory} for more intuition on the above model. For each $k$, let $\lambda_k$ denote the loss given default by obligor $k$. The total loss is therefore expressed as $L = \sum_{k=1}^K \lambda_k X_k.$
We are interested in estimating the expected loss $\Expect[L]$. Other quantities of interest include the
Value at Risk 
$\text{VaR}_\alpha := \inf\{x \mid \Prob(L \geq x) \leq \alpha\}$, and the Conditional Value at Risk $\text{CVaR}_\alpha := \Expect[L \mid L \geq \text{VaR}_\alpha(L)]$. However, we do not discuss their estimation here.

\subsection{GCI Circuit Setup}
We explain encoding the independent Gaussian model.  We use $n_z$ qubits to encode a discretized Gaussian variable $Z$, and allocate one qubit for each $X_i$. Thus the circuit will output the following quantum state:
$\ket{\Phi} = \sum_{i=1}^{2^{n_z}} \sqrt{p(z_i)}\ket{z_i} \bigotimes_{k=1}^K \sqrt{p_k(z_i)} \ket{1} + \sqrt{1-p_k(z_i)}\ket 0,$
where the discrete variable takes on values $z_1, z_2, \ldots$ with probability $p(z_1), p(z_2), \ldots$, respectively.

We set the upper and lower bounds of the discretized Gaussian, and the parameters \( p_k^0 \) and \( \rho_k \) for the 
GCI
model. The circuit can be constructed directly using the built-in class provided by Qiskit. For example, we set $n_z = 4$, and the truncation threshold between $-3$ to $3$ for the discretized Gaussian. We also follow \shortciteN{egger2020credit} and set $K = 2, p_1^0 = 0.15, p_2^0 = 0.25, \rho_1 = 0.1, \rho_2 = 0.05, \lambda_1 = 1, \lambda_2 = 2,\alpha = 0.05$ for the GCI model. The circuit can then be constructed as follows:

\begin{lstlisting}[language=Python, backgroundcolor=\color{gray!10}, basicstyle=\ttfamily\small]
import numpy as np
from qiskit.circuit.library import GaussianConditionalIndependenceModel
n_z = 4
z_max = 3
z_values = np.linspace(-z_max, z_max, 2**n_z)
p_zeros = [0.15, 0.25]
rhos = [0.1, 0.05]
lgd = [1, 2]
alpha = 0.05

#construct the circuit
GCI = GaussianConditionalIndependenceModel(n_z, z_max, p_zeros, rhos)
\end{lstlisting}

Running this circuit followed by measuring all the qubits yields a classical bitstring of length $n_z + K$, where the first $n_z$ bits represent the discretized Gaussian $Z$, and the remaining $K$ bits correspond to the realization of $(X_1, \ldots, X_K)$. The constructed circuit can be executed using the \colorbox{gray!10}{Sampler} class  in \colorbox{gray!10}{Qiskit}.

\begin{lstlisting}[language=Python, backgroundcolor=\color{gray!10}, basicstyle=\ttfamily\small]
GCI_measure = GCI.measure_all(inplace=False)
sampler = Sampler()
## executing the GCI circuit independently 2,000 times
job_1 = sampler.run(GCI_measure, shots = 2e3)
job_1.result()

\end{lstlisting}

The above code runs the GCI circuit independently 2,000 times, generating 2,000 i.i.d.\ binary vectors. As a sanity check, we set \colorbox{gray!10}{shots = 2,000} and \colorbox{gray!10}{20,000}, respectively, and plot the histogram of the discretized $Z$ in
Figure~\ref{fig3} (Left). 
As observed, the discretized  $Z$ takes values in the range $(-z_{\max},\, (-1 + 2^{-n_z + 1}) z_{\max},\, \ldots,\, z_{\max})$. As the number of samples increases, the histogram gradually approaches the shape of a bell curve.

\subsection{Amplitude Estimation in Action}\label{Subsec:Amplitud_in_Action}

To apply quantum versions of Monte Carlo for estimating statistics associated with the total loss $L$, the next step is to design two  circuits: one that encodes the loss function $L: \{0,1\}^K \rightarrow [0,\infty)$; and another that maps the loss value to the amplitude of an auxiliary qubit. The ``quantum'' way to encode such a function is using the following circuit:
 ~ $
\mathcal S: \ket{x}\ket 0_{n_S} \rightarrow \ket{x}\ket{L(x)}_{n_S}.$

Here, $\ket{x} = \ket{x_1, \ldots, x_K}$ represents a register of $K$ qubits, analogous to the classical $n$-bit input of the function $L$, and $\ket{\cdot}_{n_S}$ denotes a register of $n_S$ qubits. If $\lambda_1, \ldots, \lambda_K$ are all positive integers, then one can take $n_S = \lceil \log_2(\lambda_1 + \lambda_2 + \ldots + \lambda_K)\rceil + 1$ since all possible outcomes of \( L \) can be expressed in binary expansion.

Next, we need a circuit that maps $L$ to the amptitude of an auxiliary qubit: 
~ $\mathcal C: \ket L_{n_S} \ket 0 \rightarrow \ket L_{n_S} \left(\sqrt{\frac{L_{n_S}}{2^{n_S}- 1}}\ket 1 + \sqrt{1 -\frac{L_{n_S}}{2^{n_S}- 1}}\ket 0\right).$ Therefore, estimating $\bE[L]$ is equivalent to estimating the amptitude of the state $\ket 1$ in the final qubit
\begin{align}\label{eqn:objective qubit}
\left(\sqrt{\frac{L_{n_S}}{2^{n_S}- 1}}\ket 1 + \sqrt{1 -\frac{L_{n_S}}{2^{n_S}- 1}}\ket 0\right),
\end{align}
and then multiply by the normalizing factor $2^{n_S} - 1$. Both circuits are available in \colorbox{gray!10}{qiskit.circuit.library}.

\begin{center}
  \includegraphics[width=0.4\textwidth]{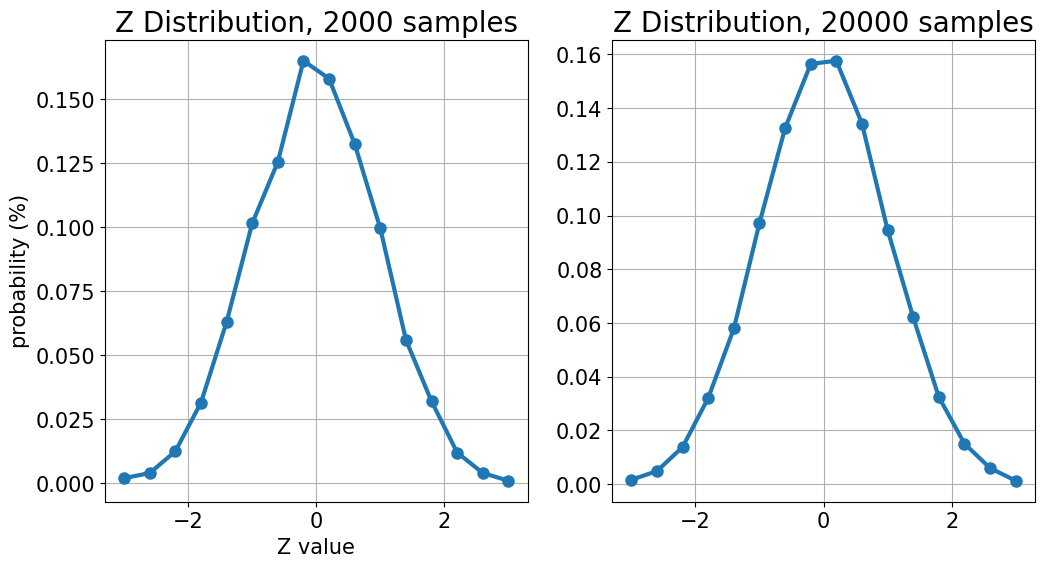}%
  \hspace{0.2\textwidth}%
  \includegraphics[width=0.24\textwidth]{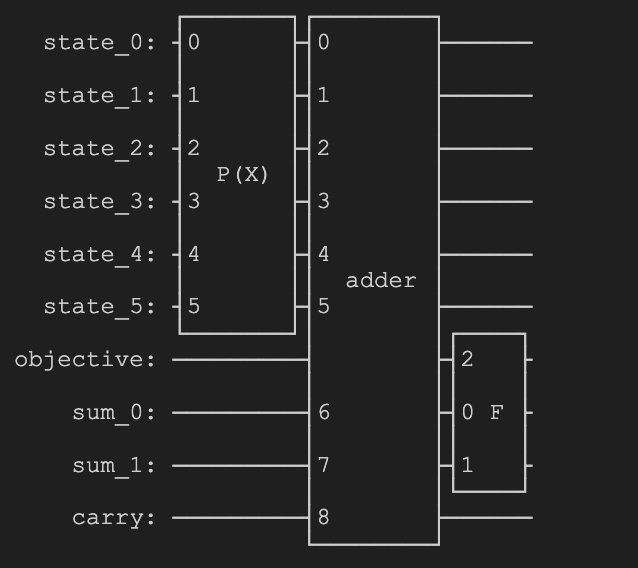}

  \vspace{0.5ex}
  \captionof{figure}{%
    Left: Histogram of Gaussian variable $Z$.
    Right: Our quantum circuit.
  }
  \label{fig3}
\end{center}

Our full  circuit has three components: we first apply the GCI circuit to encode the underlying distribution, then use the adder circuit $\mathcal{S}$ to encode the loss $L$, and finally apply the amplitude circuit $\mathcal{C}$ to encode $L$ into the amplitude. The overall circuit is their composition, $\mathcal{C} \circ \mathcal{S} \circ \text{GCI}$. We can also draw the whole circuits in qiskit; see 
Figure~\ref{fig3} (Right). 
The leftmost $P(X)$ circuit encodes the GCI model. The middle adder circuit maps the inputs $(X_1, X_2)$ to the loss value $L(X_1, X_2)$. The final circuit `F', encodes the loss $L$ into the amplitude of the objective qubit \eqref{eqn:objective qubit}, which is the $7$th qubit from top to bottom. Finally, we apply the QAE algorithm to estimate the expectation and report both the point estimate and the corresponding confidence interval.

\begin{lstlisting}[language=Python, backgroundcolor=\color{gray!10}, basicstyle=\ttfamily\small]
from qiskit_algorithms import IterativeAmplitudeEstimation, EstimationProblem
ae = IterativeAmplitudeEstimation(
    epsilon_target=epsilon, alpha=alpha, 
    sampler=RefSampler(options={"shots": 1000, "seed": 1}))
result = ae.estimate(problem)
conf_int = np.array(result.confidence_interval_processed)
print("Estimated value:\t%.4f" % result.estimation_processed)
print("Confidence interval: \t[%.4f, %.4f]" % tuple(conf_int))
\end{lstlisting}
In this example, the estimated value is $0.6681$, with a 95\% confidence interval of $[0.6435, 0.6928]$. It aligns well with the ground truth value of $0.6446$.

\section{Conclusion and Further Resources}
\label{sec:conclusion}

In this tutorial, we have set the stage by explaining why quantum computing is attracting interest and how its different architectures (gate-model vs.\ annealing) operate and require error correction. We covered the basic elements of quantum computing using Grover’s algorithm to illustrate superposition, interference, and the structure of quantum algorithms. Building on that, we explored quantum approaches to Monte Carlo methods and showed how a variation of Grover’s search (QAE) can estimate expected values with a quadratic speedup. A financial example demonstrated how
QAE
might be applied to nested expectation problems  in risk analysis or option pricing, with a sketch of a Qiskit implementation provided for clarity. 

We believe that research on quantum speedup of Monte Carlo methods is worthwhile pursuing. There are basic features that we take for granted in classical algorithms which are not as easy to translate to quantum algorithms without incurring a significant computational overhead. For instance, acceptance/rejection is a common practical and generic tool for sampling generation. Of course, there are many settings, especially in high dimensions in which acceptance/rejection is not practical and advanced sampling techniques are needed. In the quantum setting, state preparation is in general not immediate even for moderate dimensions and there is no direct general and practical tool that at least in principle yields a desired target distribution.

\section*{ACKNOWLEDGMENTS}
Blanchet, Szegedy, and Wang  acknowledge support from grant NSF-CCF-2403007 and NSF-CCF-2403008.


\section*{AUTHOR BIOGRAPHIES}

\noindent {\bf \MakeUppercase{Jose Blanchet}} is a Professor of MS\&E at Stanford University. His research interests include applied probability, Monte Carlo methods, stochastic optimization and quantum computing. His email address is \email{jose.blanchet@stanford.edu} and his website is \url{https://joseblanchet.com}.\\

\noindent {\bf \MakeUppercase{Mark S.\  Squillante}} 
is a Distinguished Research Scientist in
Mathematics of Computation 
at
IBM Research.
His research interests broadly concern 
mathematical foundations of the analysis, modeling and optimization of the design, control and performance of stochastic systems.  
His email address is \email{mss@us.ibm.com} and his website is \url{https://research.ibm.com/people/mark-squillante}.\\

\noindent {\bf \MakeUppercase{Mario Szegedy}} is a Distinguished Professor in the Department of Computer Science  at Rutgers University, New Brunswick. His research primarily focuses on quantum computing, computational complexity theory, and recently on AI. His e-mail address is \email{szegedy@cs.rutgers.edu} and his website is \url{https://people.cs.rutgers.edu/~szegedy/homepage.html}.\\

\noindent {\bf \MakeUppercase{Guanyang Wang}} is an Assistant Professor in the Department of Statistics at Rutgers University, New Brunswick. His research primarily focuses on Monte Carlo methods, generative AI, quantum computing, and probability. His e-mail address is \email{gw295@stat.rutgers.edu} and his website is \url{https://guanyangwang.github.io/}.\\

\end{document}